# *Salmonella* chemoreceptors McpB and McpC mediate a repellent response to *L*-cystine: a potential mechanism to avoid oxidative conditions

Milena D. Lazova[1§], Mitchell T. Butler[2§], Thomas S. Shimizu[1,*] and Rasika M. Harshey[2,*]

[1]FOM Institute for Atomic and Molecular Physics (AMOLF)
1098 XG Amsterdam
The Netherlands

[2]Section of Molecular Genetics and Microbiology &
Institute of Cellular and Molecular Biology
University of Texas at Austin
Austin, TX 78712

[§]These authors contributed equally to the work

*Correspondence to either author
Phone: 512-471-6881; Fax: 512-471-1218; E-mail: rasika@uts.cc.utexas.edu
Phone: 31-20-754-7242; Fax: 31-20-754-7290; E-mail: t.shimizu@amolf.nl

Running Title: McpB and McpC mediate a repellent response to *L*-cystine

**Summary**

Chemoreceptors McpB and McpC in *Salmonella enterica* have been reported to promote chemotaxis in LB motility-plate assays. Of the chemicals tested as potential effectors of these receptors, the only response was towards *L*-cysteine and its oxidized form, *L*-cystine. Although enhanced radial migration in plates suggested positive chemotaxis to both amino acids, capillary assays failed to show an attractant response to either, in cells expressing only these two chemoreceptors. *In vivo* fluorescence resonance energy transfer (FRET) measurements of kinase activity revealed that in wild-type bacteria, cysteine and cystine are chemoeffectors of opposing sign, the reduced form being a chemoattractant and the oxidized form a repellent. The attractant response to cysteine was mediated primarily by Tsr, as reported earlier for *E. coli*. The repellent response to cystine was mediated by McpB / C. Adaptive recovery upon cystine exposure required the methyl-transferase/-esterase pair, CheR / CheB, but restoration of kinase activity was never complete (*i.e.* imperfect adaptation). We provide a plausible explanation for the attractant-like responses to both cystine and cysteine in motility plates, and speculate that the opposing signs of response to this redox pair might afford *Salmonella* a mechanism to gauge and avoid oxidative environments.

## Introduction

*Salmonella enterica* and *Escherichia coli* show chemotaxis toward amino acids and sugars, as well as oxygen and other stimuli that change cellular energy levels (reviewed in (Stock & Surette, 1996, Taylor *et al.*, 1999, Alexandre & Zhulin, 2001, Hazelbauer *et al.*, 2008, Baker *et al.*, 2006, Miller *et al.*, 2009, Wadhams & Armitage, 2004)). Chemotaxis can be metabolism-independent or -dependent and requires processing of sensory input from chemoreceptors through a signaling pathway wherein the receptor-associated kinase CheA transfers phosphoryl groups to the response regulator CheY, which ultimately modulates the rotational bias of the flagellar motor. The steady-state level of the phosphorylated response regulator is determined by the balance of its production by CheA and destruction by a phosphatase CheZ. The activity of the receptor-kinase complex is feedback-regulated by the methyltransferase CheR and the methylesterase / deamidase CheB. The competing activities of CheR and CheB, involving reversible receptor methylation at multiple sites, enable cells to adapt to static chemical environments by restoring receptor-kinase output towards its pre-stimulus state.

Binding of chemoeffector molecules to transmembrane chemoreceptors, also known as methyl-accepting chemotaxis proteins (MCPs), is sufficient to initiate the metabolism-independent chemotaxis response. Ligand binding can be either direct or via a periplasmic binding protein (Neumann *et al.*, 2010). Reversible ligand binding to dimeric MCPs at their periplasmic domains affects the receptors' conformational state on both sides of the cytoplasmic membrane, thereby propagating a signal into the cell. Upon crossing the membrane, signal transmission is thought to proceed through the regulatory HAMP domain (Zhou *et al.*, 2009, Zhou *et al.*, 2011), the "methylation module" harboring the reversibly modified residues, and the signal-output domain that regulates the activity of CheA. Conserved pentapeptide motifs (NWE$^{T}$/$_{S}$F) at the C-termini of a subset of MCPs reversibly bind CheR and CheB (Barnakov *et al.*, 1999, Li & Hazelbauer, 2006). Another metabolism-independent chemotactic response involves carbohydrate transport via the phosphoenolpyruvate-dependent phosphotransferase system (PTS), which requires the CheA–CheY signaling pathway and one or more chemoreceptor species (Lux *et al.*, 1995). Metabolism-coupled chemotaxis includes redox taxis in response to changes in the redox state of the electron transport system (Bespalov *et al.*, 1996) and pH taxis in response to changes in the pH gradient across the cell membrane (proton-motive force) (Kihara & Macnab, 1981).

In *E. coli*, chemotaxis is carried out by one of four MCPs: Tsr senses serine (Mesibov & Adler, 1972), Tar senses aspartate and maltose (Mowbray & Koshland, 1987), Trg senses ribose, galactose and glucose (Kondoh *et al.*, 1979), and Tap senses dipeptides (Manson *et al.*, 1986). Trg and Tap lack the NWE$^{T}$/$_{S}$F motif and therefore require the presence of Tsr or Tar for efficient methylation-dependent adaptation to their ligands (Feng *et al.*, 1999). An additional MCP-like receptor, Aer, mediates responses to changes in oxygen concentration (Bibikov *et al.*, 1997). Aer lacks the adaptive methylation module as well as a large periplasmic domain, and it senses changes in the redox potential using a cytoplasmic PAS domain (Watts *et al.*, 2004, Bibikov *et al.*, 2004). Structural and biochemical studies indicate that chemoreceptors oligomerize as trimers of dimers, interacting at their distal cytoplasmic tips (Hazelbauer et al., 2008). The principal trimer contact residues are identical in Aer and the MCPs, suggesting that all the different receptors should be able to form mixed trimers of dimers (Gosink *et al.*, 2006). Chemoreceptors cluster in subpolar patches (Maddock & Shapiro, 1993), and there is direct experimental evidence for inter-dimer methylation (Li *et al.*, 1997).

*S. enterica* lacks Tap but has additional transmembrane chemoreceptors: Tcp that senses citrate and phenol (Yamamoto & Imae, 1993), and two recently identified receptors McpB and McpC with unknown ligand specificity (Frye *et al.*, 2006, Wang *et al.*, 2006). Two other chemoreceptor homologs with unknown function, Tip and McpA, have also been described in *S. enterica*: Tip is a transmembrane receptor with no recognizable periplasmic domain (Russo & Koshland, 1986), whereas McpA appears to be cytoplasmic (Frye et al., 2006). The *mcpC* gene is located immediately downstream of *aer*. Both genes have distinct flagellar class 3 promoters, yet insertions in *aer* are polar on *mcpC* (our unpublished results). The relative RNA levels of the *mcpB* and *mcpC* genes, as determined by microarray data, fall between those of the genes encoding the low-abundance receptor Trg and the high-abundance receptor Tsr, and are similar to the RNA levels seen for the *tar* gene (Wang et al., 2006). Both chemoreceptors have a periplasmic sensory domain, a HAMP domain, a methylation module, and receptor-trimer contact sites (Fig. S1). However, they display differences in the C-terminal pentapeptide sequence, which is NWETF in Tsr and Tar. The pentapeptide EWVSF at the C-terminus of McpB resembles NWETF at the critical positions W and F (Shiomi *et al.*, 2000), but the pentapeptide DTQPA at the C-terminus of McpC has no similarity to the NWETF sequence. In addition, the C-terminal 'tail' of McpC is 26 residues shorter than that of McpB (Fig. S1).

The present study was undertaken to identify chemoeffectors sensed by McpB and McpC, which mediate enhanced radial migration on LB or tryptone soft-agar plates (Wang et al., 2006). Here we present experimental evidence that McpB and McpC, when present as sole chemoreceptors, mediate a chemotactic response to *L*-cystine. Whereas behavior in long-time motility-plate assays shows an almost identical tactic response to both *L*-cystine and *L*-cysteine, *in vivo* fluorescence resonance energy transfer (FRET) experiments with wild-type bacteria reveal responses of opposite sign to these two chemicals that form a redox pair: cystine acts as a repellent and cysteine as an attractant. Only cystine is sensed via McpB / C. The attractant-like response to cystine in long-time behavioral assays is likely from spreading due to increased tumbling caused by a repellent response with imperfect adaptation. We discuss a possible role for the cystine response in assisting the escape of *Salmonella* from cellular damage-inducing oxidative environments.

## Results

*McpB / C mediate a response to L-cystine / L-cysteine in soft-agar assays*

In an earlier study, progressive deletion of chemoreceptors in *S. enterica* had shown that a strain missing seven of the nine chemoreceptors (Δ7T) - Tsr, Tar, Trg, Tcp, Aer, McpB and McpC - does not spread in Luria-Bertani (LB) or tryptone broth (TB) soft-agar swim plates; however, a strain missing the first five receptors but retaining McpB and McpC spreads significantly (Wang et al., 2006). We infer from these results that the two other uncharacterized receptors - Tip and McpA - do not contribute to the spreading observed in these plates. Indeed, the additional deletion of these two receptors in a Δ*tsr* Δ*tar* Δ*trg* Δ*tcp* Δ*aer* strain did not affect the migration phenotype as shown in Fig. 1 (compare BC only* to BC only; strains which still retain McpA and Tip are marked with an * hereafter; for example, the strain that contains only *mcpB* and *mcpC* chemoreceptor genes is referred to as "BC only", whereas the strain that contains only *mcpB*, *mcpC*, *tip* and *mcpA* chemoreceptor genes is referred to as "BC only*"). A strain expressing McpC alone was also capable of promoting faster spreading than the Δ7T strain, but slower than a strain expressing McpB and McpC together; a strain expressing McpB alone migrated only marginally faster than the Δ7T strain (Fig. 1A; see second row). McpC is encoded downstream of the genomic locus encoding Aer. However, Aer did not substantially affect the enhanced migration mediated by McpC (compare BC only* to C only* and C, Aer only*; see second row in A). The radial migration promoted by McpB / C was observed even when the plate was buffered to attenuate establishment of pH gradients (data not shown), suggesting that the response was to a chemical other than $H^+$.

To identify chemoeffectors, we tested the response of the BC only* strain in soft-agar swim plates containing minimal-glycerol media with mixtures of amino acids, sugars, succinate / pyruvate (labeled 'energy mix'; their metabolism creates oxygen gradients), nucleosides, and vitamins (see Experimental Procedures). Of the many potential attractants, only the commercial essential amino acid mixture 'MEM' (arginine, cystine, histidine, isoleucine, leucine, lysine, methionine, phenylalanine, threonine, tryptophan, tyrosine and valine) enhanced the migration response (Fig. 1B), whereas a non-essential amino acid mixture did not (see Experimental Procedures; essential and non-essential refer to requirement for growth of mammalian cells). To further dissect which of the MEM components triggered chemotactic spreading, we tested the individual amino acids present in the essential MEM mix. Of these, only *L*-cystine (a dimeric amino acid, formed by oxidation of two cysteine monomers covalently linked by a disulfide bridge; referred to henceforth as simply cystine), elicited a migration response (Fig. 1C). When cystine was omitted from the MEM mix, the migration rate of the BC only* strain was attenuated. In addition, we tested the reduced form, cysteine, and found that it enhances migration in a manner indistinguishable from cystine. Responses to serine and aspartate, which serve as major attractant ligands sensed by the chemoreceptors Tsr and Tar, respectively, are shown for comparison. The Δ(*mcpB mcpC*) strain (referred to as "ΔBC"), which retains seven chemoreceptors, showed the expected response to serine and aspartate (Fig. 1C; bacteria have migrated to the edge of these plates) but did not respond to cystine or cysteine in this assay. Migration responses of a strain expressing only McpC were weaker but otherwise qualitatively similar to the BC only* strain (data not shown).

*Capillary assays do not show an attractant response to cystine*

In *E. coli*, pioneering experiments by Adler and colleagues using capillary assays (Mesibov & Adler, 1972), established cysteine as an attractant sensed by Tsr, whereas cystine elicited no chemotactic response in the same assays; cysteine was also reported to be an attractant for *Salmonella* (Hedblom & Adler, 1983). Because oxidation / reduction reactions interconvert these two amino acids, it is unclear whether these redox species are stable over the many hours over which motility-plate assays are conducted, and the enhanced migration conferred by McpB and McpC could be due to cystine, cysteine, or a mixture of the two. We therefore performed Adler-type capillary assays, which are completed within a much shorter time (< 1 h), to test the response to these amino acids (Adler, 1973). We ascertained that both amino acids maintained their structure in freshly prepared solutions using mass spectrometry (see Experimental Procedures). The response of four strains – wild-type (WT), ΔBC, BC only* and Tar only* - is shown in Fig. 2, with the response to aspartic acid serving as a control (Fig. 2A). Neither BC only* nor any other of the tested strains accumulated significantly in capillaries containing cystine, indicating the lack of an attractant response (Fig. 2B). However, WT and ΔBC strains showed an attractant response to cysteine (Fig. 2C). These observations suggest that neither cysteine nor cystine is an attractant sensed by McpB / C under the conditions of these capillary assays, in stark contrast to the seemingly positive chemotactic migration response of the BC only* strain in both cysteine and cystine motility plates (Fig. 1C).

*FRET experiments reveal responses of opposite sign to the cystine / cysteine redox pair*

To probe the effect of cysteine and cystine on chemotactic activity, we used an *in vivo* fluorescence resonance energy transfer (FRET) assay, utilizing the donor-acceptor pair between fusions of CheZ and CheY to cyan and yellow fluorescent proteins (CFP and YFP), respectively (Sourjik *et al.*, 2007) (see Experimental Procedures). The FRET signal is proportional to the activity of CheA, the central kinase of the chemotaxis pathway. An analogous *in vivo* FRET system has been used in numerous studies of *E. coli* chemotactic signaling (Sourjik & Berg, 2002, Sourjik, 2004, Shimizu *et al.*, 2010, Lazova *et al.*, 2011). A schematic representation of the FRET system is shown in Fig. S2A.

We first applied step increases in the concentration of cystine to immobilized bacterial populations kept under constant flow of motility buffer, and monitored the FRET response (see Experimental Procedures). Fig. 3 (left) shows a typical time series of the FRET response to addition and removal of 100 µM cystine in *Salmonella enterica* LT2 strains. Cystine caused an increase of the FRET signal, indicating a repellent response (Fig. 3 left WT, Fig. S2B). WT responded to concentrations of cystine as low as 10 nM (data not shown). In contrast to cystine, the reduced form, cysteine, produced a decrease in the FRET signal, indicating an attractant response (Fig. 3 right WT, Fig. S2C). This response of WT cells to cysteine steps was detectable in FRET down to a threshold of ~20 µM (data not shown). The attractant response to cysteine is consistent with the capillary assay data shown in Fig. 2C as well as results from previous studies (Melton *et al.*, 1978).

In *E. coli* and *Salmonella*, efficient adaptation to chemoeffectors involves methylation and demethylation of specific glutamyl residues on the chemoreceptors by CheR and CheB respectively. In CheR / CheB+ cells (e.g. the WT strain used here), the rapid initial increase in the FRET signal upon stepping up the cystine concentration (Fig. 3 left, Fig. S2B) was followed by a slower, partial recovery toward the pre-stimulus level; upon stepping down the concentration, a small, transient decrease of the FRET signal was observed. This result showed that the repellent response to cystine was adaptive, but that the adaptation was incomplete, i.e. imperfect adaptation (Meir *et al.*, 2010, Lan *et al.*, 2011). In contrast, the FRET response of Δ(*cheR cheB*) cells to a cystine step did not recover toward the pre-stimulus level (Fig. 3 left), indicating that the adaptive recovery of the FRET response in WT cells was due to the activities of CheR and CheB. Similarly, an adaptive response to cysteine was observed in WT bacteria, and no adaptation occurred in Δ(*cheR cheB*) cells (Fig. 3 right, Fig. S2C). However, the adaptation of WT cells to cysteine was perfect: during the cysteine step, the FRET signal recovered precisely to the pre-stimulus level. Deleting the gene encoding the scaffolding protein CheV (Alexander *et al.*, 2010), whose homolog has been implicated in the chemotactic adaptation of *Bacillus subtilis* (Karatan *et al.*, 2001), showed no substantial effect on the response to cystine or cysteine (Fig. 3, bottom panels).

*The repellent response to cystine is mediated by McpB / C*

We performed FRET experiments in receptor knockout strains to probe whether cystine and cysteine are sensed in a McpB / C-dependent manner. Fig. 4 (left) shows a typical time series of the FRET response to addition and removal of 100 µM cystine in WT, ΔBC, and BC only* strains. Note that all three strains are *Salmonella enterica* 14028 derivatives, in contrast to the LT2 strains shown in Fig. 3. The differences in amplitudes in LT2 and 14028 backgrounds could be explained by the presence of unlabeled *cheY* and *cheZ* genes in 14028 strains, as well as strain-dependent variations in chemoreceptor expressions. This conjecture is supported in data presented in Figure S3 (see Experimental Procedures for details). The response to cystine was completely abolished in the ΔBC strain. However, the BC only* strain showed a repellent response to cystine: qualitatively, the temporal profile of the response was similar to WT, although the response amplitude in the BC only* strain was smaller than that of WT (Fig. 4, left), likely because of the diminished size of the total receptor population (Sourjik, 2004). Indeed, overexpression of McpC from a plasmid in the BC only strain produced a substantially stronger response (see Fig. 5A). In agreement with the capillary-assay results (Fig. 2), the FRET response of the ΔBC strain to the cysteine was nearly the same as WT, but the response to cysteine was completely abolished in the BC only* strain (Fig. 4, right). We conclude that the repellent response of *S. enterica* to cystine depends on McpB / C chemoreceptor. The reduced cysteine form is an attractant, but it is not sensed via McpB or McpC.

*Roles of McpB / C in cystine sensing and of Tsr / Tar in cysteine sensing*

We sought to dissect the roles of McpB and McpC in the cystine response by comparing FRET responses of additional mutant strains engineered for their chemoreceptor composition. For both *mcpB* and *mcpC* single-deletion strains (referred to as ΔB and ΔC respectively), the response upon cystine addition was in the repellent direction (Fig. 5A, top row), suggesting that each of these receptors can sense cystine in absence of the other. When both receptors are deleted, the response to cystine is abolished as shown on Fig 4. However, the response upon cystine removal was atypical in the ΔC strain: the FRET signal increased upon chemoeffector removal instead of decreasing, as expected for removal of a repellent. A plausible explanation for this peculiar ΔC response is that one or more of the seven other receptor species in this strain are responding to traces of cysteine present within the cystine solution (due to partial reduction of the dissolved cystine; see below). Next we probed the responses mediated by McpB and McpC when they were present in cells as the sole chemoreceptor species. Weak but detectable repellent responses to cystine were observed in the McpC only strain; the response of the McpB only strain was even weaker (Fig. 5A, middle row). Overexpression of McpC in the BC only strain produced a response comparable to wild-type; however the overexpression of McpB in the BC only strain did not noticeably increase the amplitude of the response (Fig. 5A, bottom row).

Previous studies using capillary assays demonstrated that as in *E. coli* (Mesibov & Adler, 1972) Tsr (and not Tar) is likely the dominant sensor for cysteine in *S. enterica* (Hedblom & Adler, 1983) (see also Fig. 2C). Fig. 5B shows a typical time series of the FRET response upon addition and removal of 100 µM cysteine in *S. enterica* strains deleted for the *tsr* and *tar* genes, singly and together. The response of the Δ*tar* strain was similar to wild-type; however, the amplitude of the response of the Δ*tsr* strain was strongly diminished. No attractant response to cysteine was observed in Δ(*tsr tar*) cells, even when cysteine concentrations up to 10 mM were tested (data not shown). Thus, FRET experiments confirmed the results from previous studies that Tsr is the dominant receptor for cysteine.

*Function of the C-terminal pentapeptide of McpB*

Both plate (Fig. 1A) and FRET (Fig. 5A) experiments showed that the strongest responses to cystine were observed when McpB and McpC were present together. Similar to other MCPs, both *mcpB* and *mcpC* genes have a conserved methyl-accepting domain (Fig. S1), and FRET experiments demonstrated that adaptation to cystine occurs in CheR- and CheB-dependent but CheV-independent manner (Fig. 4). This result was confirmed in both wild-type and BC only* backgrounds by motility-plate assays: deleting *cheR*, *cheB*, or *cheW* dramatically diminished migration on cystine motility plates, whereas deleting *cheV* had little effect on the cystine response (Table 2A, rows 1-10). McpB could provide 'adaptational assistance' to McpC by supplying the C-terminal pentapeptide sequence (referred to henceforth as 'pentapeptide') (Fig. S1). This sequence motif, found also at the extreme C-terminus of Tsr, Tar, and Tcp but not in the low-abundance receptors Trg and Tap, is known to stimulate the activities of CheR and CheB in *E. coli* (Barnakov et al., 1999). Low-abundance receptors mediate effective taxis only in the presence of pentapeptide-containing receptors (Feng *et al.*, 1997). Adding a flexible linker ending in the pentapeptide to the carboxyl terminus of low-abundance receptors greatly enhances their function (Weerasuriya *et al.*, 1998, Feng et al., 1999).

To test whether the weaker taxis mediated by McpC alone was due to lack of a pentapeptide sequence and whether the role of McpB was to provide this sequence, we added the last 30 residues from the C-terminus of Tsr to McpC in the C only* strain and deleted the pentapeptide from McpB in the BC only* strain. Table 2B (rows 11-14) shows a comparison of the migration of these strains in minimal-media supplemented with cystine. Addition of the Tsr

C-terminus to C only* abrogated its activity (row 13), whereas deletion of the McpB pentapeptide in BC only* resulted in spreading similar to the C only* strain (row 14). Although loss of the stimulatory effect of McpB upon deletion of its pentapeptide is consistent with a role for McpB in adaptational assistance, it could also be due to loss of McpB activity as a result of the deletion. A similar loss of activity appears to be the case with addition of the Tsr C-terminal segment to McpC.

Next, we constructed a Tar C only* strain to test if Tar could provide adaptational assistance to McpC (Table 2B, rows 15-16). This strain was efficient in its response to aspartate (row 16), but did not restore the cystine response to levels seen with the BC only* strain (compare rows 15 and 11). We also assessed the contribution of Tar and Tsr expressed from plasmids (pTar and pTsr) in the C only* strain, and compared their migration in media with cystine (Table 2B; rows 17-19) versus aspartate (Table 2B; rows 20-22) and serine (Table 2B; rows 23-25). We also introduced a plasmid pTsr$^{R64C}$ encoding Tsr with a mutation in the serine binding pocket (R64C), which cannot sense serine but is otherwise functional (Burkart *et al.*, 1998), to determine if this aided taxis of a C only* strain in LB medium (Table 2B; rows 26-28). In none of these strains did motility improve to levels seen with the BC only* strain. In summary, these data show that whereas deletion of the pentapeptide in McpB eliminates its stimulatory effect, provision of Tar or Tsr does not improve McpC-mediated taxis to cystine. Therefore, if the function of McpB is to provide adaptational assistance to McpC, then the assistance must be specific, as Tsr and Tar are unable to provide it.

## Discussion

To our knowledge, McpB / C are the first chemoreceptors reported to respond to *L*-cystine. Although the cystine response was first discovered by observing enhanced migration in motility-plate assays and interpreted as an attractant response, measurement of kinase activity using *in vivo* FRET revealed a McpB / C-specific response indicative of a repellent. Below, we tie together the apparently contradictory responses of McpB / C to cystine / cysteine in motility-plate and FRET assays.

*A unified interpretation of a repellent response to cystine*

1. Imperfect adaptation. Motility-plate assays show an apparently positive response to cystine, whereas FRET assays show a repellent response. We can reconcile the behavior in motility-plate assays by the FRET data showing 'imperfect adaptation' to cystine. CheR / B-mediated recovery does not restore kinase activity exactly to the pre-simulus level upon step stimulation with cystine (Figs. 3, 4, 5, S2A), and such imperfect adaptation could explain the enhanced spreading of cells on cystine motility plates. As was first described by Wolfe & Berg (Wolfe & Berg, 1989), radial spreading of cells on soft-agar plates can occur even in strains incapable of normal chemotaxis, e.g. in adaptation-deficient Δ(*cheR cheB*) strains, or even in "gutted" strains of *E. coli* deleted for all receptors and chemotaxis genes. In such non-chemotactic strains, the rate of spreading was found to increase monotonically with the tumbling bias. So, when a chemoeffector is seen to enhance the rate of spreading in motility plates, it could be due to an attractant response to a chemical being consumed, an increase in the steady-state tumbling bias, or both. In the context of our experiments, an increase in the steady-state tumbling bias due to imperfect adaptation to the repellent cystine would be expected to increase the rate at which cells spread in the motility-plate assays. Therefore, the imperfect adaptation to cystine observed in FRET assays forms the basis of our proposal that the enhanced migration in motility-plate assays is due to an increased rate of spreading resulting from an increase in the steady-state tumbling bias, rather than a positive chemotactic response to an attractant.

To further support this explanation, we conducted two additional short-time behavioral assays. The first was a chemical-in-plug assay first described by Tso & Adler (Tso & Adler, 1974), where bacteria are suspended uniformly at a visible turbidity in soft agar, and respond to a repellent in the plug by generating a zone of clearing around a plug within 30 minutes. This assay worked moderately well only with wild-type bacteria. Similar to that seen with the known repellent leucine (Tso & Adler, 1974), a clear zone encircled by a ring was observed around the cystine plug (Fig. S4). In addition, we monitored the motor-switching response of tethered wild-type cells and observed immediate switching to clockwise (CW) rotation of the cell body upon cystine addition, also indicating a repellent response (data not shown).

Two recent studies have provided explanations for imperfect adaptation to attractant stimuli, such as serine and aspartate (Lan et al., 2011, Meir et al., 2010). Although the details of these two proposed mechanisms differ, both are essentially due to effects of the finite number of methylation sites possessed by chemoreceptors. Whether such mechanisms might contribute to the observed imperfect adaptation to cystine would make for an interesting question for future investigations.

2. The cysteine / cysteine redox pair. We showed that cystine but not cysteine is sensed by the BC only* strain (Fig. 4). Why then do both amino acids elicit a response in motility-plate assays (Fig. 1C)? A major difference between the motility-plate, capillary and FRET assays is the time scale over which responses reveal themselves. Motility-plate assays compare colony propagation rates over hours, capillary assays reflect the accumulation of cells over minutes, and FRET assays reveal intracellular signaling responses within seconds. Because oxidation / reduction reactions interconvert cystine and cysteine, one possible explanation is that the migration response on cysteine plates is due to oxidation of cysteine to cystine during the long duration of the experiment. The similarity in the results for cystine and cysteine plates (Fig. 1C) could be explained if in both cases the cystine / cysteine ratio relaxes toward an equilibrium that is independent of the initially added form of these inter-convertible amino acids. Reports suggest that aerobic conditions would favor cystine, whereas anaerobic conditions would shift this equilibrium towards cysteine (Shinohara & Kilpatrick, 1934, Asquith & Hirst, 1969, Ehrenberg *et al.*, 1989). Indeed, when the plates were incubated anaerobically, the response of the BC only* strain to both amino acids was diminished but was again identical for the two amino acids, indicating that the equilibrium has likely shifted towards cysteine and that cystine is the true chemoeffector sensed by these receptors (Fig. 6A). (Reducing agents such as dithiothreitol or β-mercaptoethanol were not used to create reducing conditions because they are not stable for a long time in the conditions used in motility-plate assays (Stevens *et al.*, 1983)). The inference for interconversion of the cystine / cysteine redox pair was confirmed when the cysteine solution was allowed to sit at room temperature for 72 h, whereupon it generated a repellent response in the Δ(*tsr tar*) strain, which is insensitive to the cysteine (data not shown).

We also performed an alternate chemical-in-plug assay where the chemical gradient is formed by diffusion rather than consumption of the chemical. A hard-agar plug containing the chemical was inserted into a soft-agar minimal media plate and bacteria were allowed to migrate toward the plug after being inoculated at some distance (Fig. S5). This assay gave results similar to those shown in Fig. 1, confirming that McpB / C are sufficient for the taxis response to cysteine / cystine. When this assay was conducted anaerobically (Fig. 6B), the response was consistent with the results in Fig. 6A: migration towards either amino acid was not as pronounced as under aerobic conditions. Taken together, these results suggest that the equilibrium composition of the cystine / cysteine mixture shifts towards cystine under aerobic conditions and cysteine under anaerobic conditions, so that the enhanced spreading mediated by McpB / C (which senses cystine but not cysteine) is attenuated under anaerobic conditions.

*Role of McpB and McpC in the cystine response*

The strongest responses to cystine were observed when McpB and McpC were expressed together. However, because cystine responses were observed in the absence of either one, but not both of these receptors, apparently each receptor senses cystine. The requirement for the adaptation enzymes CheR and CheB was observed in both long- and short-time assays. McpB, which has the C-terminal pentapeptide motif that is absent in McpC, might provide adaptational assistance to McpC. However, because two other pentapeptide-harboring receptors, Tsr and Tar, failed to improve the function of McpC, it appears that the contribution of McpB to the McpC-mediated response is specific.

*Physiological significance of the cystine response*

Although cystine is neither a direct participant in biochemical pathways, nor incorporated into proteins, it is cystine rather than cysteine that is taken up by *E. coli* and *Salmonella* (Baptist & Kredich, 1977, Ohtsu *et al.*, 2010). At high concentrations, cysteine is toxic to cells and is exported to the periplasm by multiple cysteine transporters where it is converted into cystine in the oxidative environment of the periplasm. The periplasmic flagellar protein FliY binds cystine (Butler *et al.*, 1993) and, along with two other cystine transport systems, is implicated in its transport back into the cell (Baptist & Kredich, 1977). The cysteine / cystine shuttle system is proposed to play an important role in oxidative stress tolerance by providing reducing equivalents to the periplasm (Ohtsu et al., 2010). We have ruled out that cystine is sensed through FliY, as deletion of *fliY* in the BC-only* background did not alter its positive migration to cysteine or cystine in plates incubated under aerobic or anaerobic conditions, nor did deletion of *fliY* in the wild-type background alter the response in FRET experiments (data not shown).

What then could be the physiological relevance of the repellent response to cystine in *Salmonella*? We showed in this study that oxidized and reduced components of the cystine / cysteine redox pair elicit responses with an opposite sign: whereas cysteine is a chemoattractant, cystine acts as chemorepellent. Oxidative environments are expected to shift the equilibrium of the cysteine / cystine pair towards cystine. Therefore, the presence of cystine in the environment is likely an indicator of oxidizing conditions. Such conditions generate reactive oxygen species, which are responsible for damage to all macromolecules (DNA, lipids and proteins) (Rosner & Storz, 1997). The McpB / C-mediated repellent response to cystine could provide *S. enterica* with an escape mechanism from such environments, either outside or within the host. In oxidative environments such as those found in macrophages (McGhie *et al.*, 2009), the response to cystine could facilitate the spread of *Salmonella* beyond the gastrointestinal tract in systemic disease (Sano *et al.*, 2007).

## Experimental Procedures

*Bacterial strains, plasmids and growth conditions for motility-plate assays*

The strains and plasmids used in this study are listed in Table 1. Bacteria were grown either in L-broth (LB) base (20 g/L), tryptone broth (1% Bacto tryptone, 0.5% NaCl) or in M63 minimal medium (100 mM $KH_2PO_4$, 15 mM $(NH_4)_2SO_4$, 1.8 μM $FeSO_4.7H_2O$, 1 mM $MgSO_4$, 10 mM carbon source, adjusted to pH 7 with KOH). When testing for a response to a sugar, pre-cultures were grown with 0.2% concentration of that sugar. Amino acid (all *L*-form) and vitamin mixtures were obtained from Invitrogen, and the nucleoside mixture was purchased from Millipore. The final amino acid concentrations in minimal-swim plates ranged from 2-20 μM for individual amino acids from the Invitrogen MEM (Minimal Essential Medium) mix (arginine, cystine, histidine, isoleucine, leucine, lysine, methionine, phenylalanine, threonine, tryptophan, tyrosine and valine), or non-essential mix (glycine, alanine, asparagine, aspartic acid, glutamine, glutamic acid, proline and serine); 1 mg/L for each vitamin (choline, pantotheic acid, folic acid, nicotinamide, pyridoxal hydrochloride, riboflavin, thiamine and inositol); 30 μM for each nucleoside (adenosine, cytidine, guanosine, thymidine and uridine). Individual amino acids were tested at 100 μM. The sugar mix contained 50 μM each of glucose, maltose, ribose and arabinose. The energy mix contained 75 μM each of pyruvate and succinate. Swim or chemotaxis plates were solidified with 0.3% agar and inoculated in the center with 2.5 µl of an exponentially growing culture at $OD_{600}$ of 0.6.

Anaerobic motility assays were conducted in a 2.5 Liter, Oxoid AnaeroJar system AG0025. AnaeroGen sachets placed in a sealed jar rapidly absorb atmospheric oxygen with the simultaneous generation of carbon dioxide. Oxygen levels in the jar are claimed to fall below 1% within 30 minutes, and the resulting carbon dioxide levels are between 9% and 13%. The jar was set up according to manufacturer specifications.

*Strain and plasmid construction*

Deletion or insertion of genes and regulatory regions was achieved by the one-step mutagenesis procedure (Datsenko & Wanner, 2000) as described (Wang *et al.*, 2005). The initial deletion / substitution involved selection with either kanamycin$^{R}$ (Kan), chloramphenicol$^{R}$ (Cam) or tetracycline (Tet) cassettes. Except for deletion of the C-terminal pentapeptide encoding region of *mcpB*, all gene deletions were designed to remove the entire coding sequence except the first and last few amino acids, and were verified by DNA sequencing. LT2-based strains (TSS500, TSS507 and TSS515) were created using a modification of the Datsenko and Wanner strategy that does not leave a scar: the insertion cassette contains the lethal gene *ccdB* under control of rhamnose inducible promoter, and is removed by positive selection on rhamnose-minimal plates (Yuan & Berg, 2008). The resident plasmid pSLT (which contains *ccdA* and *ccdB* genes) was displaced prior to chromosomal manipulations using Kit 10 from *Salmonella* Genetics Stock Center (SGSC). Addition of the last 30 amino acid-encoding segment of *ts*r to the end of *mcpC* was achieved as follows: a PCR product linking the C-terminal end of *tsr* to *tet* was first generated using appropriate primers specific to *tsr* and *tet*. This product was used as a template to similarly generate a second PCR product linking the end of *mcpC* to the *tsr*-*tet* fusion. The *mcpC-tsr-tet* product was finally recombined into the C only* strain. The hybrid joint has the following sequence: DTQPA AREVAAVKTPAAVSSPKAAVADGSDNWETF, where the underlined residues are from *mcpC*, followed by those from *tsr*.

In LT2-based strains for FRET experiments shown on Figs. 3, 5B and S2, *cheY* and *cheZ* are deleted from the chromosome and CheY-YFP and CheZ-CFP fusions are expressed from a plasmid pVS88 (see Table 1). In these Δ(*cheY cheZ*) strains, lack of competitive interaction of labeled and unlabeled CheY and CheZ proteins leads to a greater amplitude of FRET responses, compared to strains that express unlabeled CheY and CheZ (such as 14028-based strains used in the experiments shown on Figs. 4 and 5A). This is supported by data in Fig. S3: the amplitudes of the initial FRET response to α-methyl-aspartate (MeAsp) (Fig. S3A) and serine (Fig. S3B) are much greater in LT2 Δ(*cheY cheZ*) than in LT2 and 14028, which both contain unlabeled *cheY* and *cheZ*. Other factors that could contribute to the different response amplitudes in LT2- and 14028-based strains are possible strain-dependent and day-to-day variations in receptor expression levels, as well as the density of cells in the area of the coverslip from which FRET signals were measured (LT2-based strains attached more efficiently than did 14028-based strains, resulting in higher experiment-to-experiment variation in fluorescence levels for the 14028-based strains).

pMB1 was constructed by PCR amplification of genomic *mcpB* using primers that included *Sph*I and *Xba*I restriction sites for ligating into the same sites on the expression vector pBAD33 (Guzman *et al.*, 1995). A similar cloning strategy was used for the plasmid for McpC expression, pML19; however, the primers contained *Sac*I and *Xba*I sites, and the expression vector was pKG110. pMK113 expressing *E. coli* Tar was a gift from Michael Manson (Texas A & M University), and pJC3 expressing *E. coli* Tsr was a gift from Sandy Parkinson (University of Utah, Salt Lake City); expression is from the *tac* promoter in the pTrc99 vector (Amann *et al.*, 1988). *E. coli* Tsr(R64C) was expressed from the parent plasmid pJC3 (Burkart et al., 1998). The FRET donor–acceptor pair - CheZ-CFP and CheY-YFP – was expressed from a plasmid pVS88 under control of an isopropyl β-D-1-thiogalactopyranoside (IPTG)–inducible promoter (Sourjik, 2004).

*Cystine preparation*

Stock solution of 100 mM cystine (*L*-cystine, Calbiochem, Cat# 2470, 99.1%, for the plate and capillary assay experiments; *L*-cystine, Sigma Aldrich, Cat# 30199, Bioultra, ≥99.5% for FRET experiments) was prepared in 1M HCl. The Calbiochem product has a certified synthetic origin. Sigma Bioultra is of animal origin; however, Calbiochem cystine, as well as two other Sigma products (Cat# C7602, 98.5-101.0% - from non-animal source, and Cat# 49603, TraceCERT® - from animal origin) were tested in FRET and qualitatively similar repellent responses were obtained (data not shown).

Working solutions were prepared in minimal-glycerol M63 medium for the plate experiments, chemotaxis buffer (CB: 1x PBS, 0.1 mM EDTA, 0.01 mM L-methionine, and 10 mM DL-lactate) for the capillary assay experiments, and motility buffer (10 mM potassium phosphate, 0.1 mM EDTA, 1 μM methionine, 10 mM lactic acid, pH 7) for FRET experiments. As the working solutions were buffered, their pH was neutral. HCl in concentrations present in the working solutions did not elicit a chemotaxis response (data not shown). Control FRET measurements with 100 μM cystine dissolved directly in motility medium without using HCl, confirmed the repellent response (data not shown).

*Mass Spectrometry*

LC-MS of cysteine and cystine was performed at the University of Texas ICMB/CRED Protein and Metabolite Analysis Facility. An electrospray ion trap mass spectrometer (LCQ, ThermoFisher, San Jose, CA) coupled with a microbore HPLC (Magic 2002, Michrom BioResources, Auburn, CA) was used to acquire spectra. Cysteine was dissolved in water and cystine was dissolved in either formic acid or hydrochloric acid aqueous solutions. The samples were analyzed immediately. 10 µl of each solution was injected into HPLC and directly infused into LCQ. Automated acquisition of full scan MS spectra was executed by Finnigan Excalibur™ software (ThermoFisher, San Jose, CA). The full scan range for MS was 50-300 Da. Each solution displayed only a single peak, corresponding to the expected mass for each amino acid (not shown).

*Chemical-in-plug assays*

Two variations of this assay originally described by Tso & Adler were performed (Tso & Adler, 1974). In the long-time assays, hard-agar plugs with 10 mM chemical dissolved in minimal-glucose media and set with 2% agar were inserted with a sterile pipette tip into soft-agar (0.3%) plates made with minimal media. The plates were poured at least 5 h before use and the plugs were inserted just before the plates were point-inoculated with bacteria at some distance from the plug. Plates were incubated for >20 h at 37$^{o}$C. In the short-time assays bacteria sufficiently concentrated to give visible turbidity were uniformly suspended in soft-agar plates (~4 x$10^{9}$ cells/plate). As before, a plug of hard agar containing the chemical repellent, prepared as described by (Tso & Adler, 1974) was inserted into the soft-agar plate and monitored within 30 min at room temperature.

*Capillary Assays*

Capillary assays were performed as previously described (Adler, 1973), except that plastic gaskets (2 cm in diameter, ~1.5 mm thick) were used to create the chamber or "pond". About one sixth ($60^o$) of the circular gasket was removed to provide a portal for entry of the capillary tubes. Capillaries contained either chemotaxis buffer (CB) alone or CB with the indicated concentration of aspartate, cysteine or cystine. The first two amino acids were dissolved in deionized water, whereas cystine was first dissolved in 0.1M HCl and then neutralized with NaOH. Freshly prepared 100 mM stock solutions were diluted appropriately in CB prior to the capillary assay, which was run for 45 min at $37^oC$. The number of cells entering the capillary was determined by plating dilutions of the capillary contents on LB agar and counting colonies after 24 h incubation at $37^oC$.

*In vivo fluorescence resonance energy transfer (FRET) assay of kinase CheA activity*

The FRET pair, in which the response regulator, CheY, and its phosphatase, CheZ, are genetically fused to yellow (acceptor) and cyan (donor) fluorescence proteins (YFP and CFP) respectively, provides a measure of the concentration of the intracellular complex, formed between phosphorylated CheY (CheY-P) and CheZ. The concentration of CheZ·CheY-P complex is determined by two opposing reactions: phosphorylation of CheY by CheA, and dephosphorylation of CheY-P by CheZ. The rates of the two reactions are equal at steady-state, therefore the FRET signal is proportional to the activity of the central kinase of the chemotaxis pathway CheA, which is considered as a single output of the chemoreceptor activity (Tu *et al.*, 2008, Sourjik & Berg, 2002, Shimizu et al., 2010) (see Fig. S2A). Thus, this FRET pair provides real-time readout of the activity of the bacterial chemotaxis pathway for any changes on a time scale greater than the relaxation time of CheY phosphorylation cycle.

*FRET experiments and data analysis*

Bacteria were grown at 33.5°C to mid exponential phase ($OD_{600}$ ~0.5) in tryptone broth supplemented with appropriate antibiotics and inducers (see Table 1). Cells were harvested by centrifugation, washed twice, resuspended in motility buffer and stored at 4°C.

Prior to the experiment (1-5 h after harvesting), bacteria were immobilized on a poly-L-lysine coated coverslip. The coverslip was then situated at the top face of a flow cell (Berg & Block, 1984), and the bacteria were kept under constant flow of motility buffer generated by a syringe pump (Harvard Apparatus, PHD2000). The same flow was used to add and remove chemoeffectors during experiments. There is a consistent ~ 25 s delay between the time when the switch was thrown to induce the step (indicated by arrows on the figures) and the time when the new solution reached the cells located in the flow cell.

An upright microscope (Nikon FN1), equipped with an oil immersion objective (Nikon CFI Plan Fluor, 40x/1.3), was used to perform FRET microscopy. The sample, situated in the flow cell, was illuminated by a metal halide arc lamp with closed-loop feedback (EXFO X-Cite *exacte*) through an excitation bandpass filter (Semrock, FF01-438/24-25) and a dichroic mirror (Semrock, FF458-Di01). The epifluorescent emission was split by a second dichroic mirror (Semrock, FF509-FDi01) into donor (cyan, *C*) and acceptor (yellow, *Y*) channels. Photon-counting photomultipliers (Hamamatsu H7422P-40) were used to collect the signal from *C* and *Y* channels through emission bandpass filters (Semrock FF01-483/32 for C channel and FF01-542/27 for *Y* channel). Signal intensities of the donor and acceptor channels were recorded through a data acquisition card (National Instruments) installed on a PC running custom-written software.

Both *Y* and *C* channels were corrected for the coverslip background. The signal from the *Y* channel was also corrected for leakage from CFP emission. The ratio *R* between the two channels, $R=Y/C$, serves as a robust indicator of FRET activity. $\Delta FRET$, the change in FRET efficiency upon stimulation at every time point, was computed as following:

$$\Delta FRET = (R_{pre}+\Delta R-R_0)/(R_{pre}+\Delta R+|\Delta Y/\Delta C|) - (R_{pre}-R_0)/(R_{pre}+|\Delta Y/\Delta C|),$$

where $R_0$ is the acceptor to donor ratio in absence of FRET, $R_{pre}$ is the prestimulus accetor to donor ratio, $\Delta R=R-R_{pre}$ is the change in the ratio upon stimulation, and $|\Delta Y/\Delta C|$ is the constant absolute ratio between the changes in the acceptor and donor signals per FRET pair (Sourjik et al., 2007). Under the applied experimental conditions, $R_{pre} +|\Delta Y/\Delta C| >> \Delta R$; thus $\Delta FRET \sim\Delta R$. Thus for simplicity $\Delta FRET$ is expressed in arbitrary units of $\Delta R$.

## Acknowledgments

We thank Susana Mariconda, Jaemin Lee and Vince Nieto for strain construction, Arul Jayaraman and Manjunath Hedge for help with capillary assays, Vince Nieto for tethering experiments, and Sandy Parkinson and Mike Manson for comments on an earlier version of the manuscript. This work was supported by an NIH grant to RMH (GM 57400), and The Netherlands Organization for Scientific Research/Foundation for Fundamental Research on Matter (M.D.L. and T.S.S.).

## Figure and Table Legends

**Fig. 1**. Response of wild-type and mutant *Salmonella* strains to chemoeffectors in soft-agar swim plates. (A) Top row: WT (14028), BC only* (JW20), B only* (MB2), C only* (MB1) C / Aer only* (SM576), Δ7T* (SM162) and BC only (MB203) strains were inoculated at the center of LB swim plates and incubated at 37°C for 7 h, when the wild-type had just reached the edge. * indicates that McpA and Tip are still present in the strains. Second row: Plates shown in the top row were incubated for an additional 16 h at room temperature, by which time the BC only* strains colonized half the plate. (B) The BC only* strain was inoculated in minimal-glycerol swim media containing either the commercial MEM essential amino acid mix or the indicated nutrient mixes and incubated at 37°C for 22 h (see Experimental Procedures). (C) BC only* and ΔBC strains (SM542) were inoculated in minimal-glucose swim media containing the essential MEM mix or indicated amino acids and incubated at 37°C for 22 h. The MEM mixes in this experiment were reconstructed to reflect the composition of the commercial mix.

**Fig. 2**. Quantification of the cystine response with capillary assays. The response of WT (14028), BC only* (JW20), ΔBC (SM542) and Tar only* (SM469) strains was monitored with (A) aspartic acid, (B) cystine, and (C) cysteine. The cell numbers are an average of three technical repeats of the experiment. Error bars are standard deviation from the mean. See Experimental Procedures for assay conditions.

**Fig. 3.** FRET response of WT and mutant *Salmonella* LT2 strains to cystine and cysteine. Strains used were: WT (TSS500), Δ(*cheR cheB*) (TSS507), Δ*cheV* (TSS515). 100 µM cystine or cysteine steps were used in all panels except for Δ(*cheR cheB*) (right), where the cysteine step size was 100 mM. Up / down arrows indicate the time of addition / removal of chemoeffectors, respectively.

**Fig. 4**. FRET response of WT and mutant *Salmonella* 14028 strains to cystine and cysteine. Strains used were: WT (14028), ΔBC (SM542), BC only* (JW20). 100 µM cystine or cysteine steps were used in all panels, except for ΔBC (right), where the cysteine step was 1 mM. Insets with magnified axes are shown for strains with a weaker or no response. Three repeats are averaged for BC only* cystine and cysteine responses and for ΔBC cystine responses. Other descriptions as in Fig. 3.

**Fig. 5.** FRET responses to cystine and cysteine steps in various chemoreceptor mutants. (A) 100 µM cystine steps were used in all panels. Strains used were: ΔB (QW265), ΔC (SM457), Δ9T + C (MB211 + pML19), Δ9T + B (MB211 + pMB1), BC only + C (MB203 + pML19), BC only + B (MB203 + pMB1). McpB expression from pMB1 was induced with 0.2% *L*-arabinose, and McpC expression from pML19 was induced with 7 µM sodium salicylate. Three repeats are averaged for all. (B) 100 µM cysteine steps were used in all pretreatments. Strains used were: Δ*tar* (TSS878), Δ*tsr* (TSS868), Δ(*tsr tar*) (TSS866). Other descriptions as in Figs. 3 and 4.

**Fig. 6**. Response to cystine / cysteine in aerobic versus anaerobic conditions in long-time motility-plate assays. (A) Bacteria were inoculated in the center of minimal media plates. Growth conditions were as described in Fig. 1C. See Experimental Procedures for description of the anaerobic chamber. (B) Chemical-in-plug assay**.** The hard-agar plugs on the right contain the test chemical, which diffuses into the soft-agar media (see Experimental Procedures). Bacteria were inoculated at some distance and plates incubated for 22 h at 37°C. BC only* (JW20).

**Table 1.** Strains and plasmids used in this study.

[a]Δ and :: refer to deletion of, or deletion / substitution within the indicated gene, respectively. Kan, Cam or Tet refer to substitutions with kanamycin, chloramphenicol, and tetracycline-resistance cassettes. Note that deletions created by the Datsenko & Wanner method (Datsenko & Wanner, 2000) leave behind a 'scar' sequence of ~80 base pairs in all 14028 strains. LT2 based deletion strains do not have a scar (see Experimental Procedures). Strains indicating 'only' refer to presence only of the indicated chemoreceptor gene and absence of all others. Only* indicates that the strain retains *mcpA* and *tip*. *mcpC::tsr* strain fuses the end of *mcpC* to the C-terminal 30 amino acid residues of *tsr* and has a Tet marker downstream. *mcpB*$_{\Delta5}$ deletes of the last five amino acid residues in *mcpB.*

[b]*Salmonella* Genetic Stock Center

**Table 2**. Motility-plate chemotaxis assays.

(A) McpB / C function requires CheB, CheR and CheW but not CheV. Motility is expressed as relative swim colony diameter compared to wild-type (given an arbitrary value of 10) whose moving front had just reached the edge of an LB swim plate (37°C, 7 h; see Fig. 1A). Strains used in this assay: Δ*cheB* (SM387), Δ*cheR* (SM399), Δ*cheW* (SM464), Δ*cheV* (SM423), BC only* (JW20), BC only* Δ*cheB* (MB82), BC only* Δ*cheR* (MB83), BC only* Δ*cheW* (MB84), BC only* Δ*cheV* (MB187). * indicates that these strains retain McpA and Tip. (B) Role of the C-terminal pentapeptide NWE$^{T}/_{S}$F in McpBC function. Strains used are: BC only* (JW20), C only* (MB1), C::Tsr only* (ST1000), $B_{\Delta 5}$ C only* (ST 1001), Tar C only* (ST 998). Strains were inoculated either in LB or in minimal-swim media and incubated as indicated. 1 µM IPTG was included in the plates containing pTsr (pJC3) and pTsr$^{R64C}$ (pJC3 derivative with a T156P mutation in *tsr*; Tsr(R64C). pTar (pMK113) expresses Tar constitutively.

## Supporting Information

**Fig. S1**. Alignment of known chemoreceptors of *S. enterica* with McpB and McpC. Identified ligand-binding residues are highlighted within ovals, whereas shared homologous regions with distinct functions are color-coded, their approximate boundaries indicated with jagged edges. Not shown are Aer and Tip, which do not have a substantial periplasmic domain. The cytoplasmic receptor McpA is also not shown.

**Fig. S2**. Description of the FRET system and typical responses to cystine and cysteine. (A) Schematic representation of the FRET system used in this study (see text for details). (B) and (C) From top to bottom: changes in the yellow (*Y*) fluorescence channel, cyan (*C*) fluorescence channel, *Y*/*C* ratio and *Y*/*C* ratio corrected for baseline drift: (B) at 0 s 200 µM cystine is added, removed after 500 s; (C) at 0 s 200 µM cysteine is added, removed after 400s. (B) and (C) serve as an illustration of attractant and repellent responses of wild-type *S. enterica* (TSS500). $\Delta FRET$ in Figs. 3, 4, and 5 is plotted after baseline correction and expressed in arbitrary units of $\Delta Y/C$.

**Fig. S3.** Comparison of differences in the FRET responses to α-methyl-aspartate (MeAsp) (A) and serine (B) in the presence and absence of the native (and hence unlabeled) *cheY* and *cheZ* genes. FRET responses of LT2 Δ(*cheY cheZ*) (TSS500, referred to as LT2 YZ-), LT2 (referred to as LT2 YZ+) and 14028 strain to 100 µM MeAsp and 100 µM serine are shown. The LT2 YZ-strain shows a greater amplitude of the response to both chemoeffectors than the other two strains. Other descriptions as in Fig. 3.

**Fig. S4.** Short-time chemical-in-plug assay. In this assay, wild-type *S. enterica* (14028) were suspended uniformly at a high cell density in a soft-agar plate. Hard-agar plugs containing the test chemical were inserted in the agar, and the response monitored within 30 min at room temperature. See Experimental Procedures for more details. (A) control with no chemical added, (B) 1 mM leucine, (C) 0.3 mM cystine. Only low concentrations of cystine could be used in this assay (100-300 μM) because achieving higher concentrations requires dissolution in HCl, which by itself gives a repellent response in this short-time assay.

**Fig. S5.** Chemical-in-plug assays**.** The experimental set up was as described in Fig. 6B. At 10 mM, cystine stays soluble only in an acidic solution (0.37% HCl), so HCl controls were also included. Bacteria were inoculated away from the plug, and their migration was observed after 20 h at 37°C. BC only* (JW20).

Fig. 1

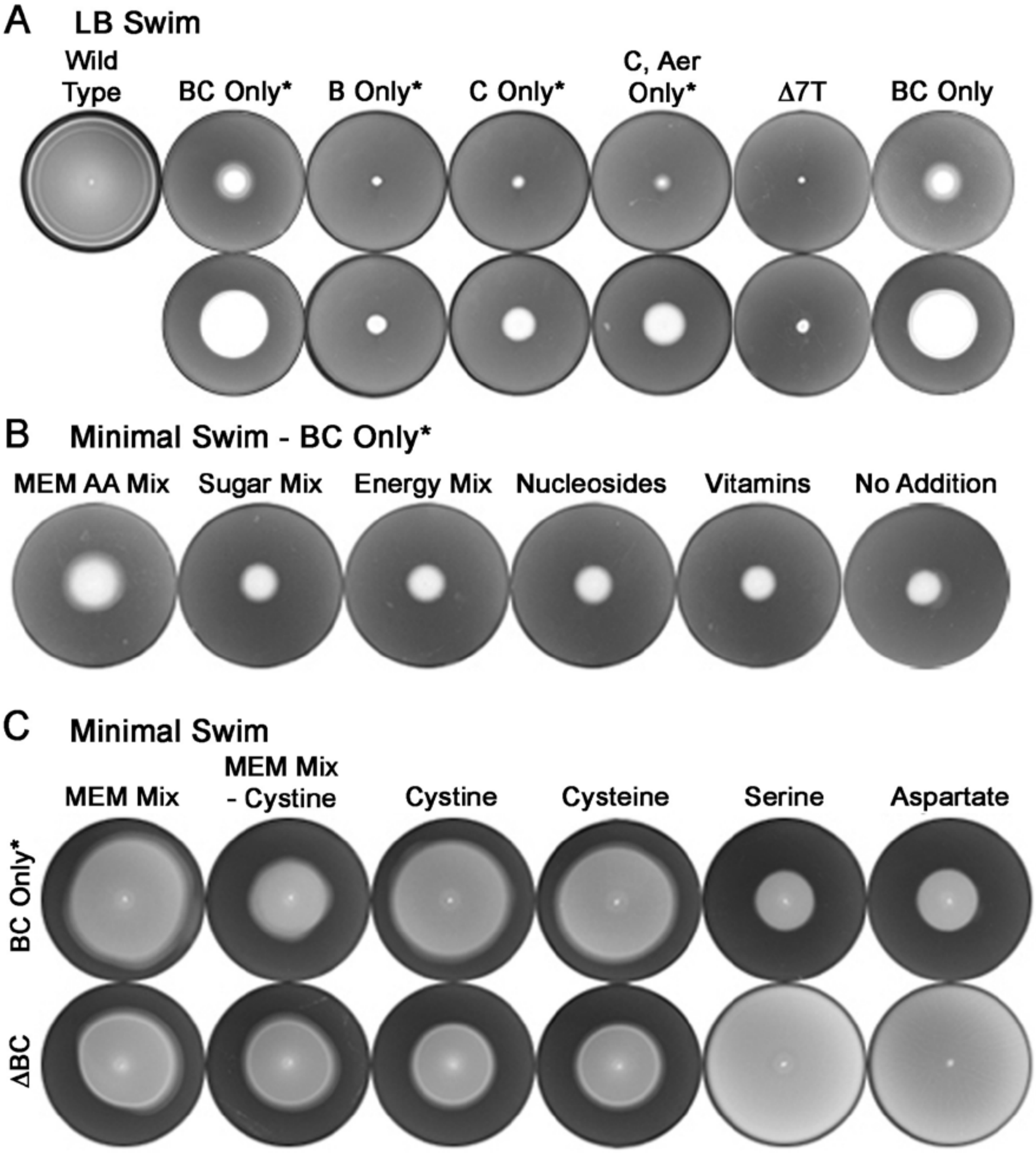

Fig. 2

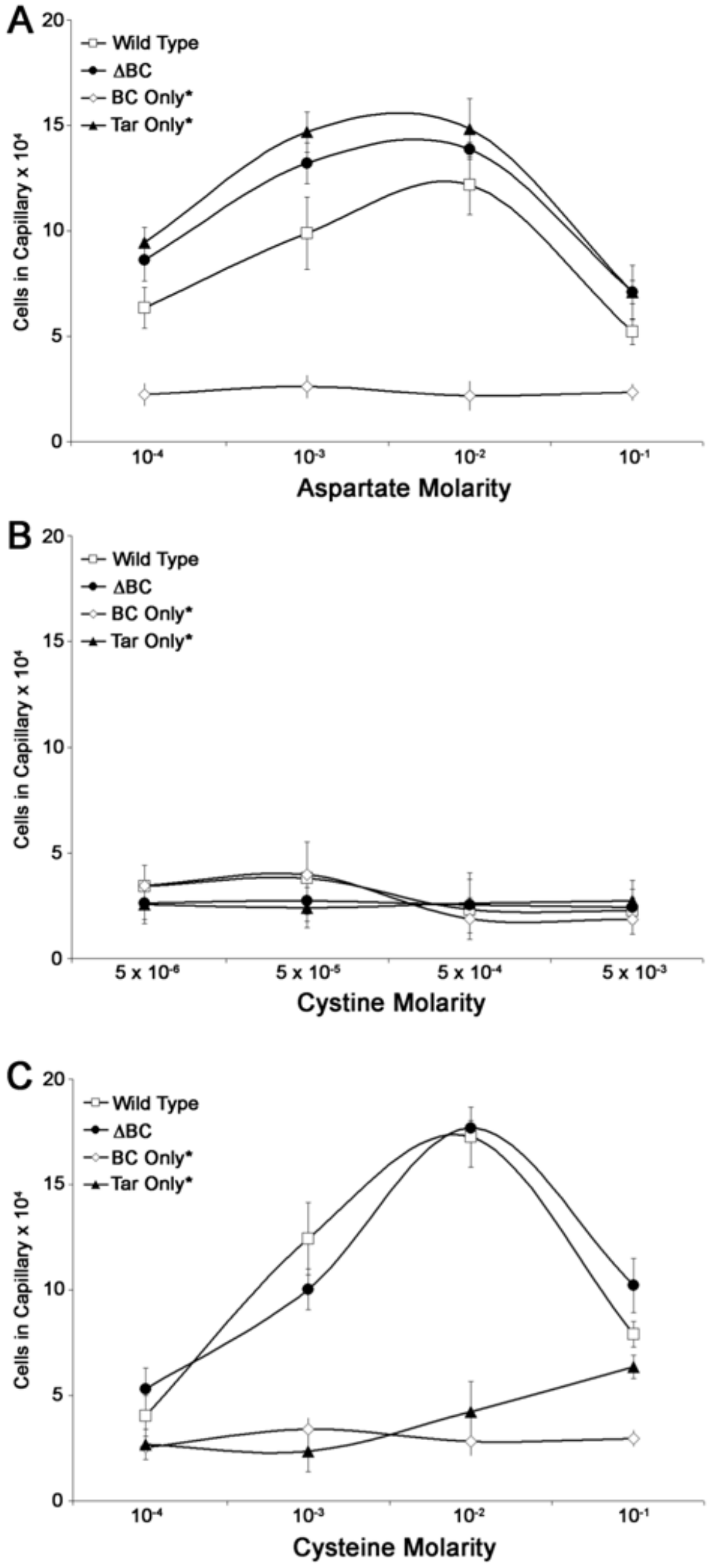

Fig. 3

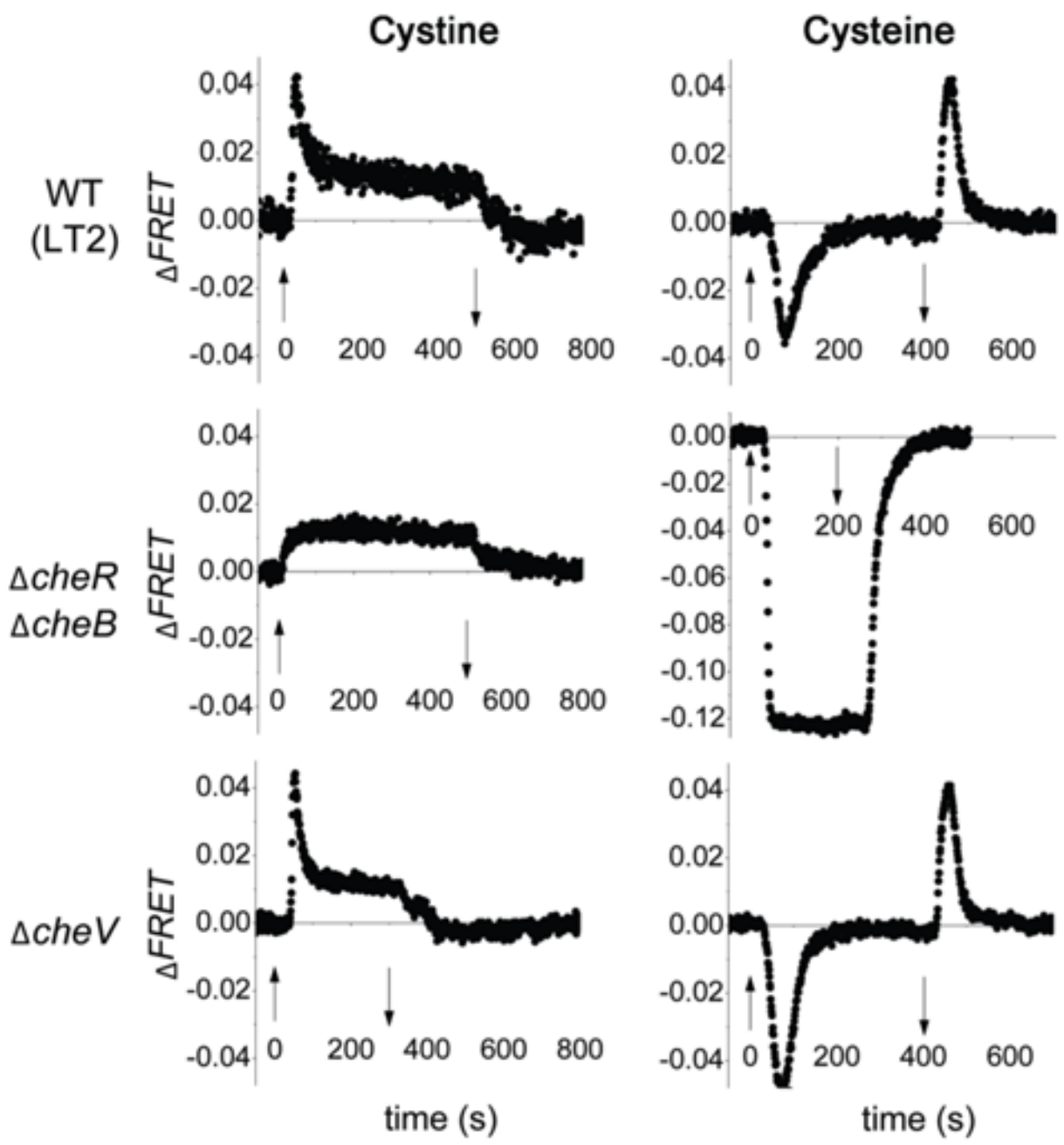

Fig. 4

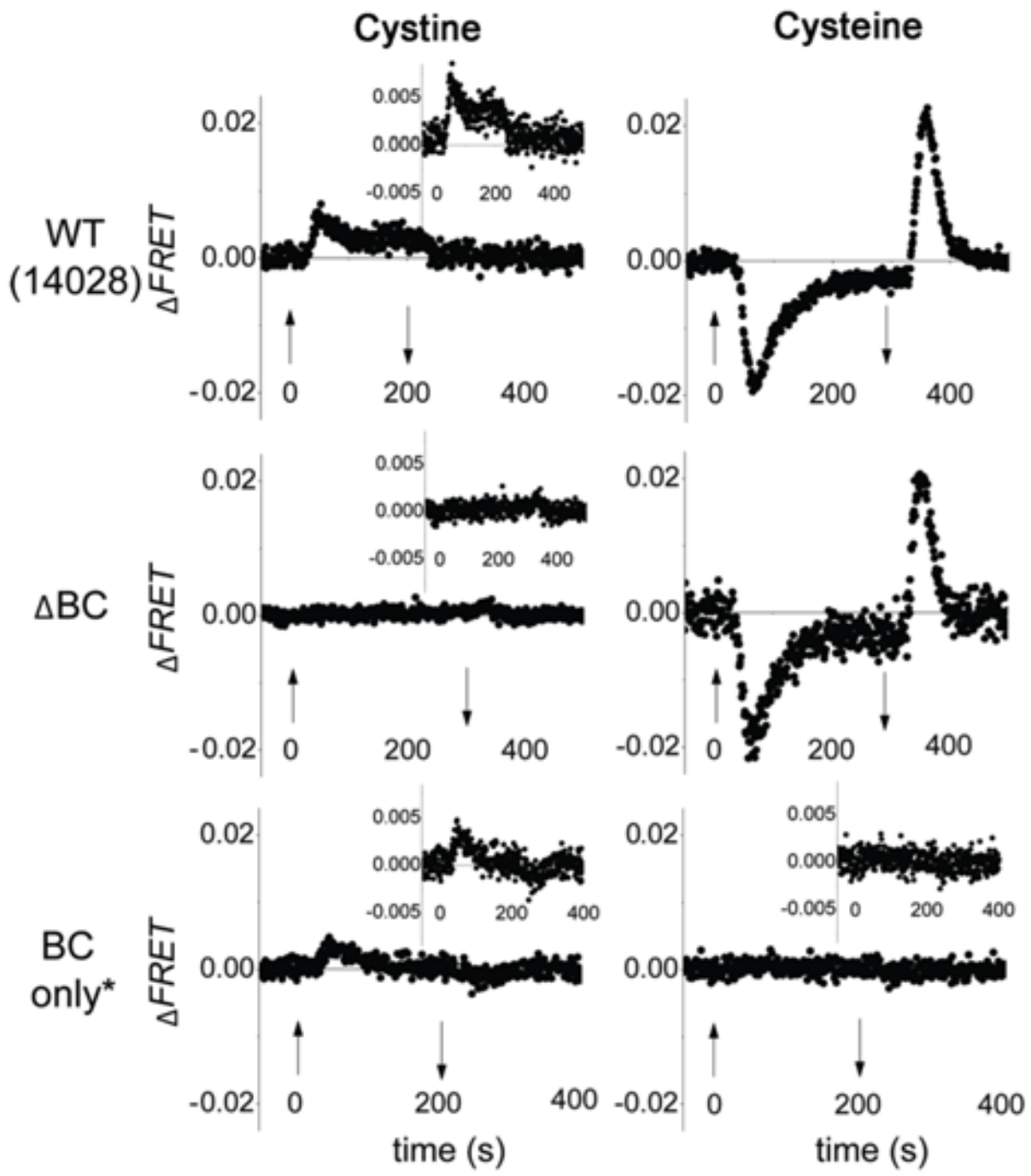

Fig. 5

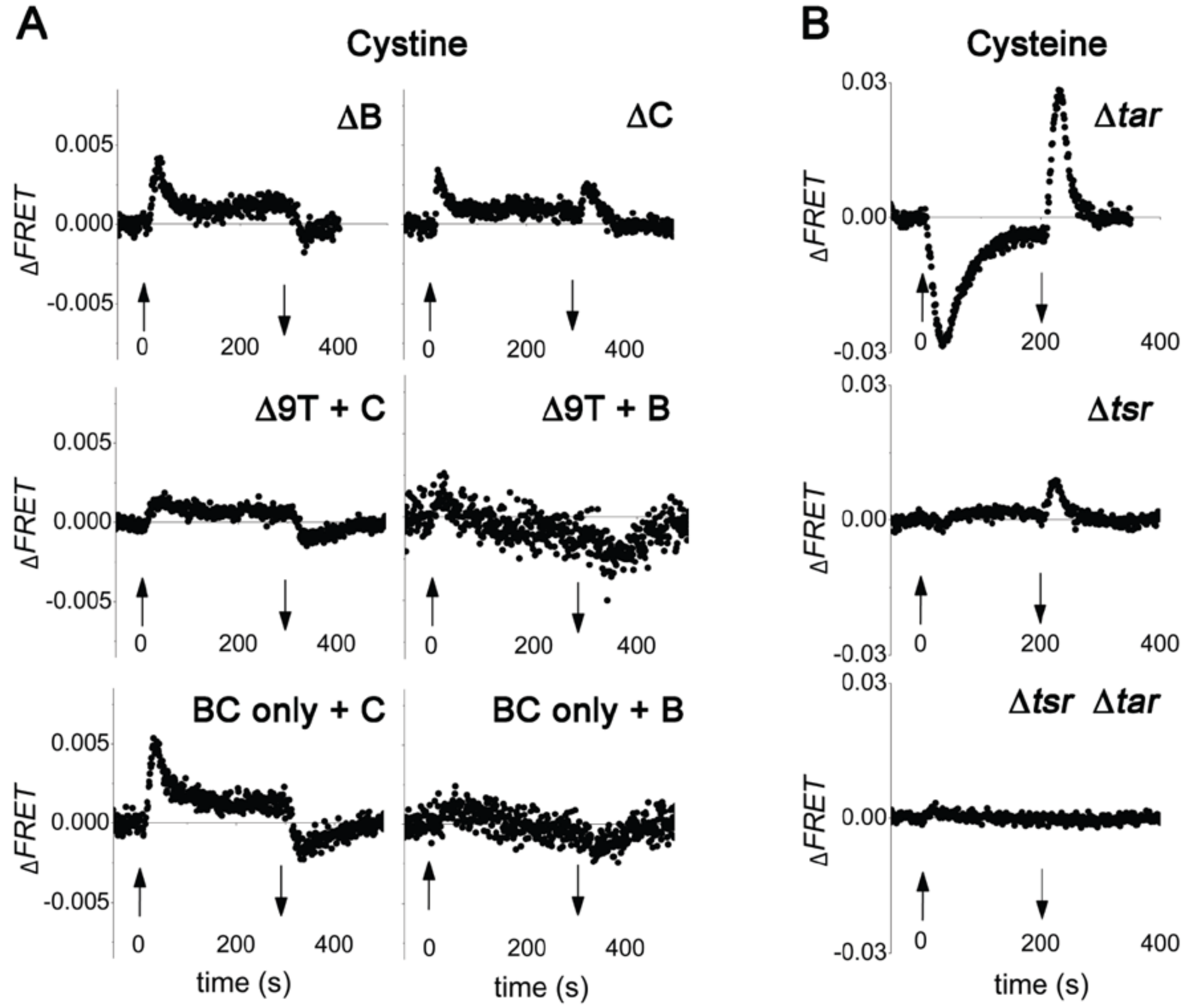

Fig. 6

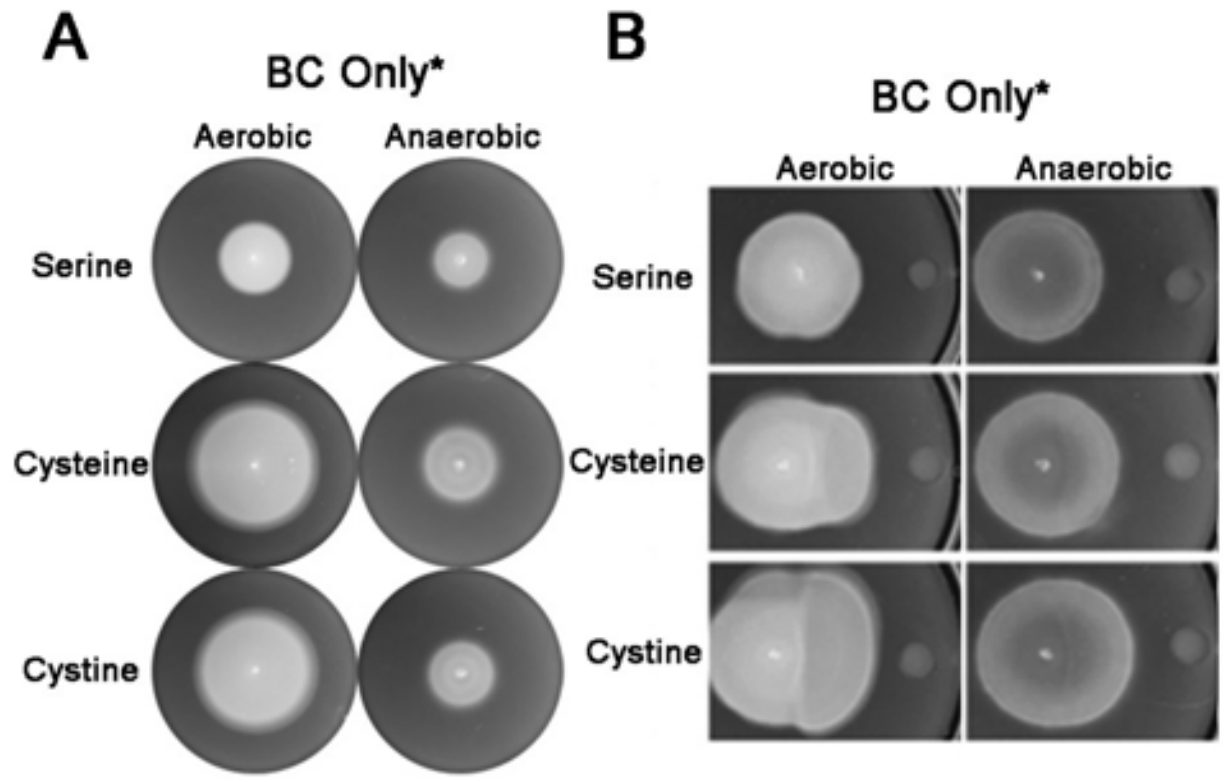

TABLE 1.

| Strain | Relevant Genotype[a] | Source/reference |
|---|---|---|
| 14028 | Wild type ATCC strain of *S. enterica* serovar Typhimurium | SGSC[b] |
| SM542 | 14028 Δ*mcpB*, Δ*mcpC* | This study |
| QW265 | 14028 Δ*mcpB* | This study |
| SM457 | 14028 Δ*mcpC* | This study |
| JW20 | 14028 *mcpB*, *mcpC* only* | This study |
| MB203 | 14028 *mcpB*, *mcpC* only (*mcpA*::Kan) | This study |
| MB1 | 14028 *mcpC* only* (*mcpB*::Cm) | This study |
| MB2 | 14028 *mcpB* only* (*mcpC*::Cm) | This study |
| MB211 | 14028 Δ*9T* (Δ*tsr* Δ*tsr* Δ*trg* Δ*tcp* Δ*aer* Δ*mcpA* Δ*mcpB* Δ*mcpC* Δ*tip*) | This study |
| SM469 | 14028 *tar* only* (*tcp*::Kan) | This study |
| SM576 | 14028 *mcpC, aer* only* (*mcpB*::Cm, *tcp*::Kan) | This study |
| SM387 | 14028 Δ*cheB* | This study |
| SM399 | 14028 Δ*cheR* | This study |
| SM423 | 14028 Δ*cheV*::Cm | This study |
| SM464 | 14028 Δ*cheW* | This study |
| MB82 | 14028 *mcpB, mcpC* only*, *cheB*::Kan | This study |
| MB83 | 14028 *mcpB, mcpC* only*, *cheR*::Kan | This study |
| MB84 | 14028 *mcpB, mcpC* only*, *cheW*::Kan | This study |
| MB187 | 14028 *mcpB, mcpC* only*, *cheV*::Kan | This study |
| SM162 | 14028 Δ*7T* (Δ*tsr* Δ*tar* Δ*trg* Δ*aer* Δ*mcpB* Δ*mcpC* *tcp::*Kan) | This study |
| ST998 | 14028 *tar mcpC* only* (*mcpB*::Cm) | This study |
| ST1000 | 14028 *mcpC::tsr*_Tet only* (*mcpB*::Cm) | This study |
| ST1001 | 14028 *mcpB*$_{\Delta 5}$, *mcpC* only* | This study |
| LT2 | *S. enterica* serovar Typhimurium str. LT2 | SGSC[b] |
| TSS500 | LT2 Δ*cheY* Δ*cheZ* | This study |
| TSS507 | LT2 Δ*cheR* Δ*cheB* Δ*cheY* Δ*cheZ* | This study |
| TSS515 | LT2 Δ*cheV* Δ*cheY* Δ*cheZ* | This study |
| TSS868 | LT2 Δ*tsr* Δ*cheY* Δ*cheZ* | This study |
| TSS878 | LT2 Δ*tar* Δ*cheY* Δ*cheZ* | This study |
| TSS866 | LT2 Δ*tsr* Δ*tar* Δ*cheY* Δ*cheZ* | This study |

| Plasmid | Gene(s) | Resistance | Replication Origin | Induction | Source/ reference |
|---|---|---|---|---|---|
| pKG110 | Cloning Vector | Chloramphenicol | pACYC | Sodium salicylate | J.S. Parkinson |
| pBAD33 | Cloning Vector | Chloramphenicol | pACYC | Arabinose | Guzman et al 1995 |
| pTrc99 | Cloning Vector | Ampicillin | pBR | IPTG | Amann et al 1988 |
| pML19 | LT2 *mcpC* | Chloramphenicol | pACYC | Sodium salycilate | This study |
| pMB1 | *mcpB* | Chloramphenicol | pACYC | Arabinose | This study |
| pVS88 | *cheZ-ecfp / cheY-eypf* | Ampicillin | pBR | IPTG | Sourjik & Berg, 2004 |
| pMK113 | *E.coli tar* | Ampicillin | pBR | Constitutive | M. D. Manson |
| pJC3 | *E.coli tsr* | Ampicillin | pBR | IPTG | J. S. Parkinson |
| pJC3 (R64C) | *E.coli tsr* insensitive to serine | Ampicillin | pBR | IPTG | Burkart et al, 1998 |

TABLE 2.

| # | Strain | Incubation Time | Media | Motility |
|---|---|---|---|---|
| **A** | Wild Type | | | |
| 1 | 14028 | 7 h at 37°C | LB Swim | 10 |
| 2 | Δ*cheB* | 7 h at 37°C + O/N at RT | " | 1 |
| 3 | Δ*cheR* | " | " | 0 |
| 4 | Δ*cheW* | " | " | 0 |
| 5 | Δ*cheV* | 7 h at 37°C | " | 9 |
| 6 | BC only* | " | " | 7 |
| 7 | BC only*, Δ*cheB* | 7 h at 37°C + O/N at RT | " | 1 |
| 8 | BC only*, Δ*cheR* | " | " | 0 |
| 9 | BC only*, Δ*cheW* | " | " | 0 |
| 10 | BC only*, Δ*cheV* | 7 h at 37°C | " | 7 |
| **B** | | | | |
| 11 | BC only* | 22 h at 37°C | Minimal Swim + Cystine | 9 |
| 12 | C only* | " | " | 4 |
| 13 | C::Tsr only* | " | " | 0 |
| 14 | $B_{\Delta 5}$ C only* | " | " | 4 |
| 15 | Tar, C only* | " | " | 5 |
| 16 | Tar, C only* | " | Minimal Swim + Aspartate | 10 |
| 17 | C only* | 22 h at 37°C | Minimal Swim + Cystine | 4 |
| 18 | C only*, pTar | " | " | 4 |
| | C only*, pTsr | " | " | 3 |
| 19 | | | | |
| 20 | C only* | “ | Minimal Swim + Aspartate | 2 |
| 21 | C only*, pTar | “ | “ | 6 |
| 22 | C only*, pTsr | “ | “ | 3 |
| 23 | C only* | “ | Minimal Swim + Serine | 2 |
| 24 | C only*, pTar | “ | “ | 4 |
| 25 | C only*, pTsr | ” | ” | 8 |
| 26 | BC only* | 7 h at 37°C + O/N at RT | LB swim | 7 |
| 27 | C only* | “ | “ | 3 |
| 28 | C only*, $pTsr^{R64C}$ | “ | “ | 4 |

Fig. S1

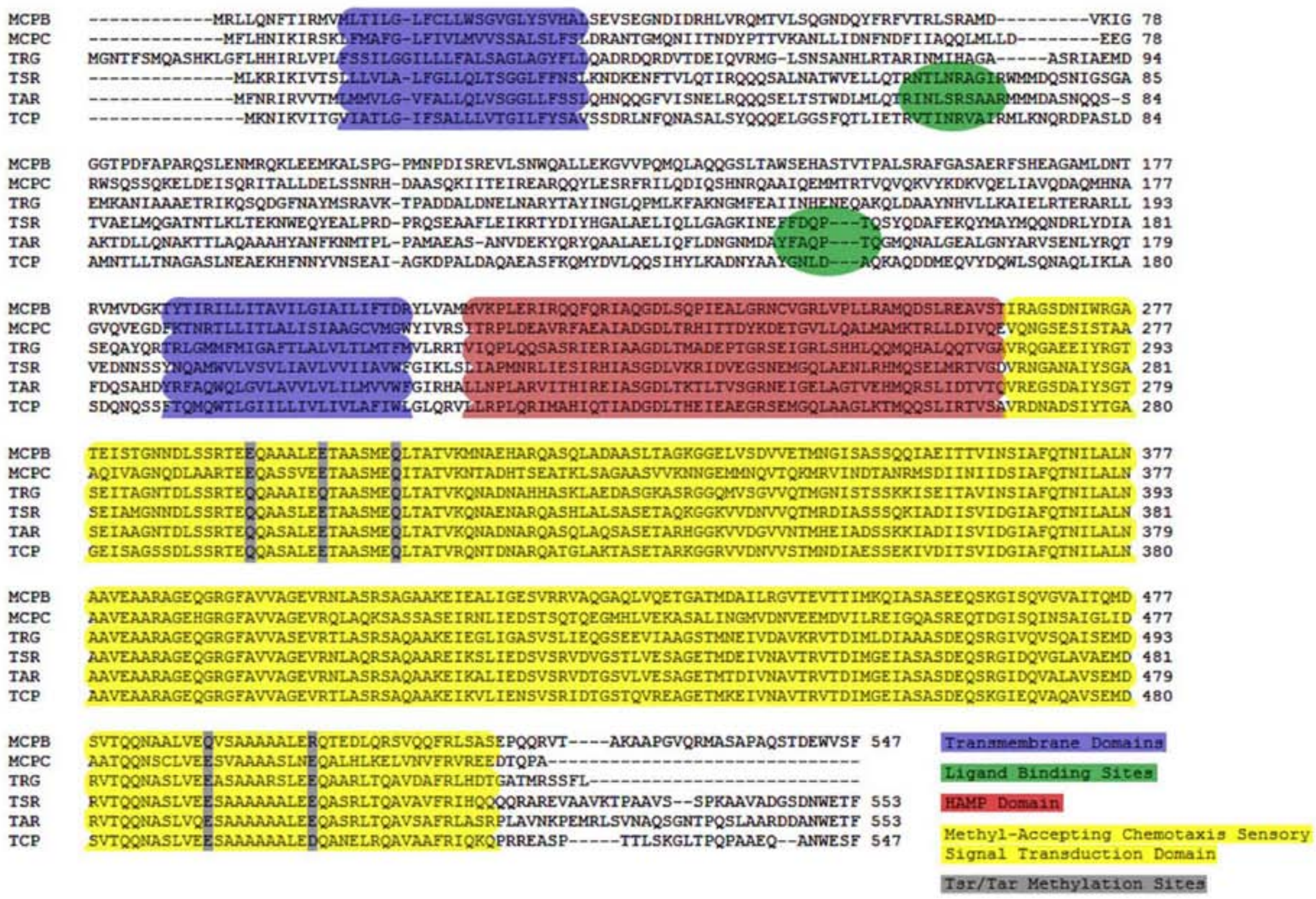
Transmembrane Domains
Ligand Binding Sites
HAMP Domain
Methyl-Accepting Chemotaxis Sensory Signal Transduction Domain
Tsr/Tar Methylation Sites

Fig. S2

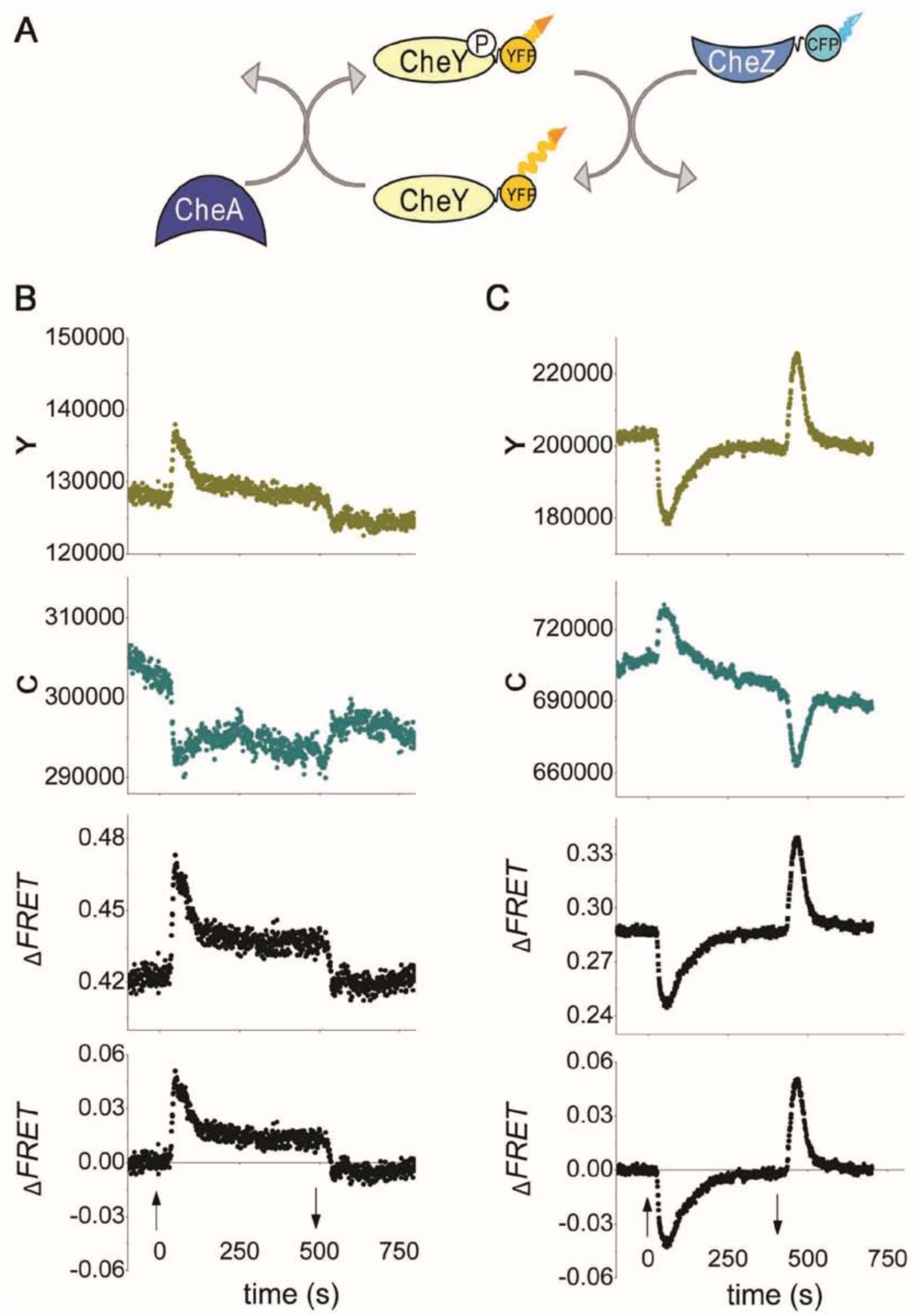

Fig. S3

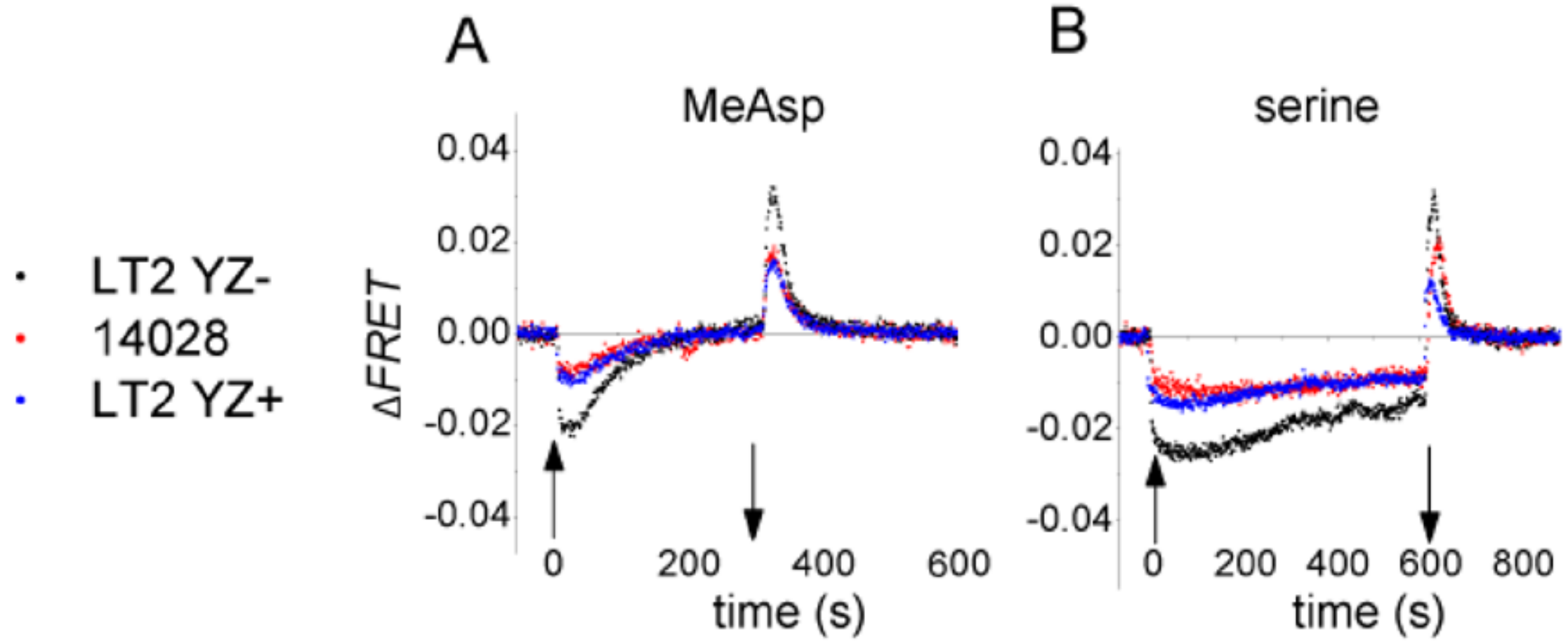

Fig. S4

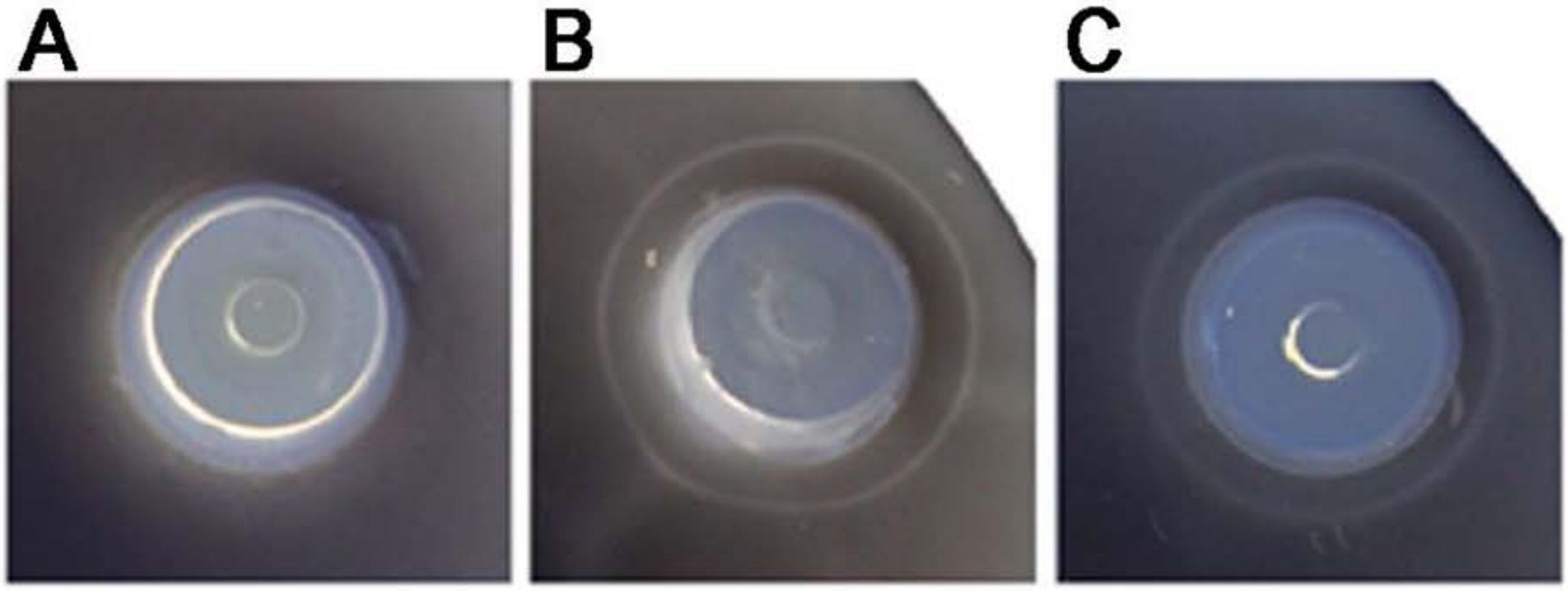

Fig. S5

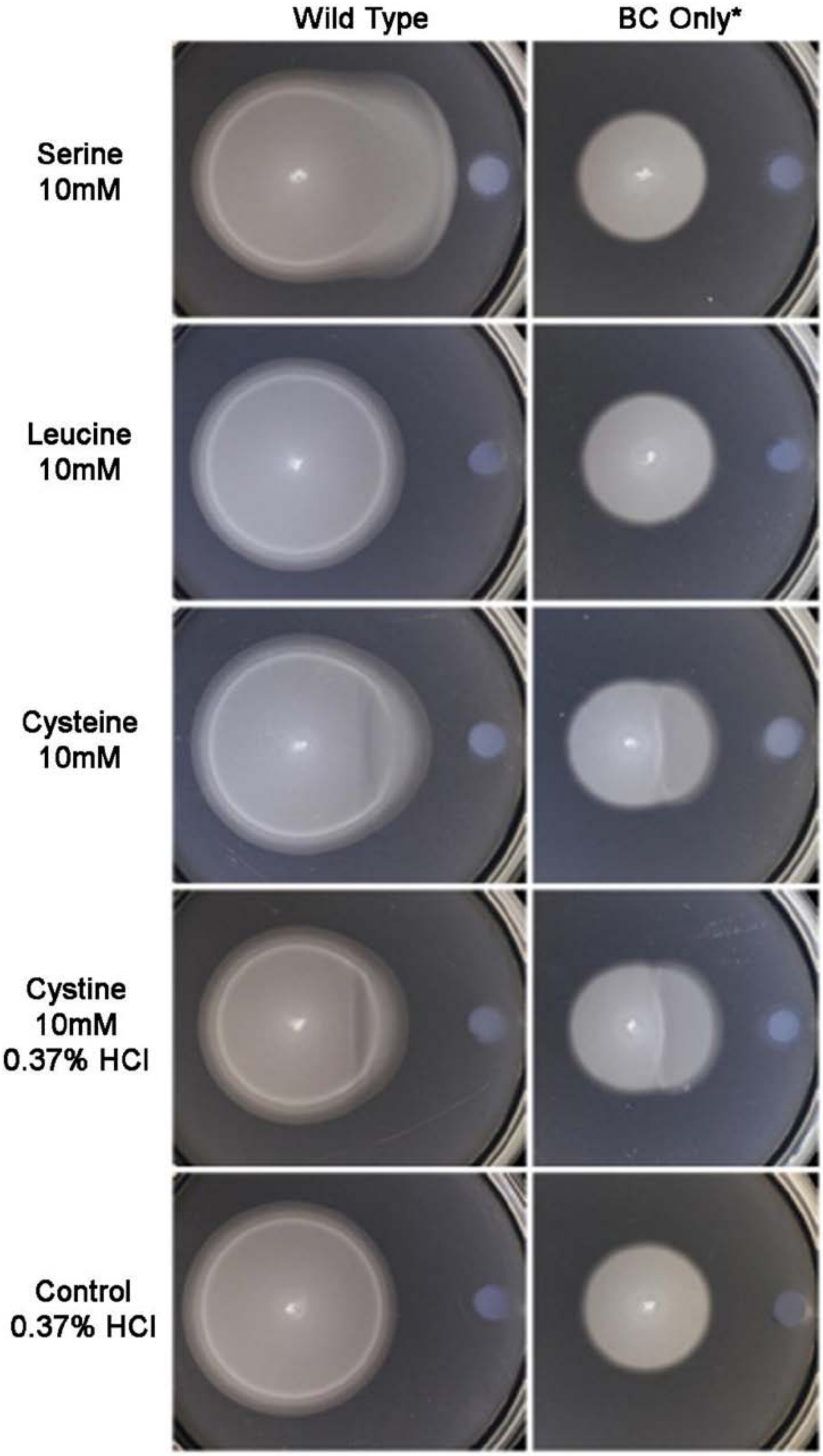